\documentclass[conference]{IEEEtran}

\usepackage{cite}
\usepackage{amsmath,amssymb,amsfonts}
\usepackage{algorithmic}
\usepackage{graphicx}
\usepackage{textcomp}
\usepackage[table]{xcolor}
\usepackage[hyphens]{url}
\usepackage{soul}
\usepackage{hyperref}

\usepackage{annotate-equations}
\usepackage{listings}
\usepackage{xspace}
\usepackage{ctable}
\usepackage{tcolorbox}
\usepackage{dirtytalk}

\usepackage{todonotes}
\usepackage[ruled,linesnumbered]{algorithm2e}
\usepackage{multirow}

\def\BibTeX{{\rm B\kern-.05em{\sc i\kern-.025em b}\kern-.08em
    T\kern-.1667em\lower.7ex\hbox{E}\kern-.125emX}}

\definecolor{color_1}{HTML}{009E73} %
\definecolor{color_2}{HTML}{D55E00} %
\definecolor{color_3}{HTML}{0072B2} %
\definecolor{color_4}{HTML}{000000} %
\definecolor{color_5}{HTML}{E69F00} %
\definecolor{color_6}{HTML}{CC79A7} %

\colorlet{soul_1}{color_1!20}
\colorlet{soul_2}{color_2!20}
\colorlet{soul_3}{color_3!20}
\colorlet{soul_4}{color_4!20}
\colorlet{soul_5}{color_5!20}
\colorlet{soul_6}{color_6!20}

\colorlet{local_load}{soul_5}
\colorlet{startup}{soul_6}
\colorlet{visits}{soul_3}
\colorlet{inters}{soul_2}
\colorlet{eod}{soul_1}

\newcommand{\highlight}[2]{\sethlcolor{#1}\hl{#2}}

\DeclareRobustCommand{\hlvisits}[1]{\sethlcolor{visits}\hl{#1}}
\DeclareRobustCommand{\hlinters}[1]{\sethlcolor{inters}\hl{#1}}
\DeclareRobustCommand{\hleod}[1]{\sethlcolor{eod}\hl{#1}}

\soulregister{\highlight}{2}

\SetKwProg{ParFor}{foreach}{ pardo}{end}

\SetKwIF{HlIf}{HlElseIf}{HlElse}{\hl{if}}{ \hl{then}}{\hl{else if}}{\hl{else}}{\hl{end}}
\SetKwProg{HlIf}{\hl{if}}{\hl{ then}}{\hl{end}}
\SetKwProg{HlFor}{\hl{for}}{\hl{ pardo}}{\hl{end}}
\SetKwProg{HlForEach}{\hl{foreach}}{\hl{ do}}{\hl{end}}
\SetKwProg{HlParFor}{\hl{foreach}}{\hl{ do}}{\hl{end}}

\SetKwFunction{KwComputeExposures}{ComputeExposures}
\SetKwFunction{KwContactProbability}{ContactProbability}
\SetKwFunction{KwIsInfected}{IsInfected}

\newcommand{\tweakedsim}{\raise.17ex\hbox{$\scriptstyle\mathtt{\sim}$}}

\definecolor{dkgreen}{RGB}{0,64,0}
\definecolor{ltgray}{RGB}{245,245,245}
\definecolor{mauve}{RGB}{139,0,139}

\begin{document}

\title{Pandemics \emph{in silico}: Scaling Agent-based Simulations
on Realistic Social Contact Networks}

\author{
  \IEEEauthorblockN{
    Joy Kitson\IEEEauthorrefmark{2},
    Ian Costello\IEEEauthorrefmark{1},
    Jiangzhuo Chen\IEEEauthorrefmark{3},
    Diego Jim\'enez\IEEEauthorrefmark{4},
    Stefan Hoops\IEEEauthorrefmark{3},
    Henning Mortveit\IEEEauthorrefmark{3},\\
    Esteban Meneses\IEEEauthorrefmark{5},
    Jae-Seung Yeom\IEEEauthorrefmark{6},
    Madhav V.~Marathe\IEEEauthorrefmark{3},
    Abhinav Bhatele\IEEEauthorrefmark{2}
  }
  \IEEEauthorblockA{~\\
    \textit{\IEEEauthorrefmark{2}Department of Computer Science, University of Maryland}, College Park, USA\\
    \textit{\IEEEauthorrefmark{1}Google, Inc.}, Mountain View, USA\\
    \textit{\IEEEauthorrefmark{3}Biocomplexity Institute and Initiative, University of Virginia}, Charlottesville, USA\\
    \textit{\IEEEauthorrefmark{4}Max Planck Computing and Data Facility}, Garching, Germany\\
    \textit{\IEEEauthorrefmark{5}National Advanced Computing Collaboratory, National High Technology Center}, San Jos\'e, Costa Rica\\
    \textit{\IEEEauthorrefmark{6}Center for Applied Scientific Computing, Lawrence Livermore National Laboratory}, Livermore, USA\\
    E-mail: \IEEEauthorrefmark{2}jkitson@umd.edu, \IEEEauthorrefmark{2}bhatele@cs.umd.edu
  }
}

\maketitle

\begin{abstract}
Preventing the spread of infectious diseases requires implementing
interventions at various levels of government and evaluating the potential
impact and efficacy of those preemptive measures. Agent-based modeling can be
used for detailed studies of the spread of such diseases in the presence of
possible interventions. The computational cost of modeling epidemic diffusion
through large social contact networks necessitates the use of parallel
algorithms and resources in order to achieve quick turnaround times. In this
work, we present Loimos, a scalable parallel framework for simulating epidemic
diffusion.  Loimos uses a hybrid of time-stepping and discrete event simulation
to model disease spread, and is implemented on top of Charm++, an asynchronous,
many-task runtime that enables over-decomposition and adaptive overlap of
computation and communication.  We demonstrate that Loimos is able to
achieve significant speedups while scaling to large core counts. In particular,
Loimos is able to simulate 200 days of a COVID-19 outbreak on a digital twin of
California in about 42 seconds, for an average of 4.6 billion traversed edges
per second (TEPS), using 4096 cores on Perlmutter at NERSC.

\end{abstract}

\begin{IEEEkeywords}
  high performance computing,
  agent-based modeling,
  epidemiology,
  social network graphs
\end{IEEEkeywords}

\section{Introduction}
\label{sec:intro}
The COVID-19 pandemic has demonstrated that while we have made significant
progress in controlling infectious disease outbreaks, such outbreaks will
continue to pose a threat.  Computational models played a critical role during
the COVID-19 pandemic in various response efforts -- to forecast the trajectory
of the pandemic, evaluate various what-if scenarios, and support economic and
logistical planning problems such as vaccine allocation and
distribution~\cite{Cramer2021-hub-dataset, cdc-covid19-smh, vdh:modeling,
uta:modeling}.
Several challenges have also emerged as a result of these efforts, including:
(1) running these models in real time, (2) scaling models to larger regions and
incorporating a range of social, behavioral, economic and immunological
considerations, and (3) managing limited data and the resulting uncertainty
regarding conditions on the ground.

Traditional modeling techniques for the spread of infectious diseases often
rely on coupled rate equations -- systems of differential equations relating
the number or proportions of people who are, for example, susceptible, exposed,
infected, and recovered (SEIR)~\cite{hethcote2000mathematics}. While such approaches
are effective at capturing \emph{statistical} trends such as the rate at which people
are infected, they smooth over much of the complexity of human social
networks and the interactions through which diseases spread. Since
interventions to mitigate or stop epidemic spread often change this network of interactions, their
impact on a disease's spread can only be modeled indirectly under this
paradigm.

In contrast, agent-based models simulate the epidemic process on social contact
networks that capture the dynamics of human interactions. While more flexible,
this approach is much more computationally expensive, requiring agent-based
models to be parallel and highly scalable.  The first reason for this is that
it is important to be able to simulate epidemic dynamics over national and
global scale networks. A realistic social contact network for the U.S.~would
have \tweakedsim 340 million agents (as of 2025), and a global scale network would have
\tweakedsim 8 billion agents.  Second, interventions are an important component
of any epidemic simulation that seeks to study the impact of government
planning and response. However, interventions complicate interaction networks
that are already highly irregular by allowing them to change over time, which
can slow simulations down considerably.

The stochasticity of this class of agent-based models also poses challenges
when it comes to evaluating a scientific workload.
Complex experiments that study several possible scenarios require many runs,
whether for sensitivity analysis, uncertainty analysis, or comparing model
projections for a wide range of different scenarios. A typical design with 20
scenarios, each with 100 slight perturbations on model parameters for
uncertainty analysis, each with 30 replicates to account for the stochasticity
of the model, can yield 60,000 simulation experiments. Performing this many
experiments in a short amount of time requires a highly scalable code. Running
such large workflows means pushing the limits of performance, and motivates the
development of a parallel program capable of scaling to meet these demands.

\begin{table*}[t]
  \centering
  \caption{Summary of results for prior agent-based epidemic simulators. *EpiHiper result is an ensemble spanning two systems.}
  \label{tab:related}
  \begin{tabular}{lrrlrrr}
    \toprule
    Simulator & No.~of Agents & No.~of Days Simulated & Machine & No.~of Cores & Execution Time & Execution Time per Day \\ \midrule
    FRED~\cite{grefenstette2013fred} & 289 million & Unknown & Blacklight at PSC & 16 & 4 h & Unknown \\
    EpiCast~\cite{germann_mitigation_2006} & 281 million & 180 & 2.4 GHz Intel Xeon & 256 & 8-12 h & 160 s \\
    EpiHiper*~\cite{bhattacharya_data-driven_2023} & 288 million & 72,000 & \begin{tabular}[c]{@{}l@{}}Bridges-2 at PSC,\\ Rivanna at UVA\end{tabular} & \begin{tabular}[c]{@{}l@{}} 6,400 \\ 1,200 \end{tabular} & 32 h 42 m & 0.141 s \\
    EpiSimdemics~\cite{bhatele2017massively} & 280.4 million & 180 & Blue Waters at NCSA & 655,360 & 10.41 s & 0.0578 s \\
    \bottomrule
  \end{tabular}
\end{table*}

Designing and implementing parallel simulations for contagion modeling is
challenging for two main reasons: (1) the underlying social contact networks on
which infectious diseases spread are highly
unstructured (see~\cite{bhattacharya2021ai, bhattacharya2022data} for an in-depth
discussion), and (2) the dynamics over such networks are stochastic in nature;
the nodes that participate in the spreading process may differ. This complicates
partitioning and load balancing, as one
cannot predict the inter-process communication and workload on each process
\emph{a priori}.

Our primary objective in this work is to develop a scalable, parallel
simulation framework for modeling contagion processes over large relational and
time-varying networked systems.  Toward this end, we present Loimos, a highly
scalable parallel code for agent-based simulations on realistic social contact
networks, written on top of Charm++, an asynchronous, many-task runtime
system.  Loimos utilizes a combination of discrete event simulation (DES) and
time-stepping to model the spread of diseases on these networks.

\vspace{0.05in}
\noindent The key contributions of this work are summarized below:
\begin{itemize}
  \item Designing and implementing a parallel
  agent-based simulator for modeling contagion processes and intervention
  scenarios at an individual level.
  \item Validating the simulator against EpiHiper~\cite{machi_scalable_2021},
  an existing model used by the CDC COVID-19 scenario modeling
  hub~\cite{truelove_projected_2022}.
  \item Identifying three major performance bottlenecks in Loimos
  and introducing optimizations addressing them.
  \item Evaluating the scalability of Loimos on the CPU partition of Perlmutter, a production supercomputer at NERSC/LBL
  in strong and weak scaling scenarios.
\end{itemize}

\section{Related Work}
\label{sec:related}
Bissett et al.~\cite{bissett_agent-based_2021} identify five components of
agent-based techniques for modeling epidemics: (1) a theory component,
(2) synthetic population construction, (3) social contact network generation
from such synthetic populations, (4) construction of idealized social contact
networks, and (5) simulation of epidemic diffusion across both types of
contact networks. We focus on this last component, as the other components
generally represent one-time costs for an outbreak response.

Several recent publications focus on modeling the spread of COVID-19.
Many of these are national or regional compartmental models built using
data from outbreaks in the simulated area, and account for interventions in
different ways. The SIQR (Susceptible Infectious Quarantined Recovered)~
\cite{tiwari2020modelling} and SIDARTHE (Susceptible Infected Diagnosed Ailing
Recognized Threatened Healed and Extinct)~\cite{giordano2020modelling} models
have been used to simulate the progression of the pandemic in India and Italy,
respectively, using new disease states and adjustments to the values of disease
parameters to capture the impact of interventions. An age-segmented SIRD
(Susceptible Infectious Recovered Dead) model using synthetic contact matrices
for interventions~\cite{prem2020effect}, and a SIRD model using an optimization
algorithm to estimate the infection rate based on empirical data~\cite{anastassopoulou2020data}
have also been used to model COVID-19. Metapopulation models are commonly used
to capture international disease spread. These models segment
the simulated population into subpopulations representing
countries or regions and build a compartmental model for each subpopulation
with flows connecting them. One such model has been used to estimate the impact of
travel restrictions on the early spread of COVID-19~\cite{chinazzi2020effect}.
There are also some efforts to build small-scale agent-based models to
simulate the spread of the virus within small communities or within single
buildings. Although these models range in complexity --
COVID-ABS~\cite{silva2020covid} incorporates both economic and
epidemiological models within a single
simulation whereas Cuevas's model~\cite{cuevas2020agent} of spread within a
building only requires two rules to guide its agents'
behavior -- most of these models are quite small, only
simulating a few hundred agents.

Several parallel agent-based epidemic simulators have been developed for
HPC systems, including several that operate on national scales. However, any
performance comparison between existing models is hampered by the lack of
detailed information on the parameters and HPC systems used in the runs. No
single ground truth dataset exists to test raw computational speed in this
domain, so we instead seek to compare simulations that operate on a similar
scale, namely that of the population of the United States. Note that most prior
work considers 280-290 million agent populations to be U.S.-scale, as
shown in Table~\ref{tab:related}.

With this limitation in mind, there are a number of approaches to developing
high-performance agent-based disease simulations.  The Framework for
Replication of Epidemiological Dynamics (FRED)~\cite{grefenstette2013fred}
is an OpenMP based simulation that uses U.S. census data to model disease
spread. FRED's disease models are fixed to a
configurable SEIR model (susceptible, exposed, infectious, and recovered),
resulting in much less flexibility in terms of input, compared to codes that
support a tunable arbitrary disease model, such as Loimos. Seal et
al.~\cite{seal:simutools2010} implement an agent-based model involving a
generalization of Conway's Game of Life, instead of dynamics on realistic
contact networks. However, this simulation is notable for its use of GPU
offloading, which most of the simulations surveyed -- along with Loimos -- do
not support. Germann et al.~\cite{germann_mitigation_2006} develop EpiCast
by adapting the SPaSM molecular dynamics simulator, using cells as an analog
for communities and particles as an analog for individual agents. EpiCast lies halfway
between a meta-population model, having spatially distributed interaction groups,
and a fully-fledged agent-based model, placing each agent in multiple
interaction groups at once, representing where they live, work, and travel.
Parker et al.~\cite{parker2011distributed} introduce a novel approach that
models how human behavior changes due to a pandemic (e.g. increased social
distancing) but is limited to an SEIR model and requires creating new populations
to model different sets of behaviors. Machin et
al.~\cite{machi_scalable_2021} and the EpiHiper team~\cite{bhattacharya2022data}
present an agent-based simulator embedded in an end-to-end pipeline that
runs the gauntlet from model calibration to simulation output analysis.  While
this represents a production workflow used by the CDC COVID-19
scenario modeling hub~\cite{truelove_projected_2022}, their work focuses more
on the orchestration of the overall pipeline than the optimization of
individual application runs. %

Perumalla et al.~\cite{perumalla2012discrete} and the EpiSimdemics team
\cite{barrett2008episimdemics,yeom2014overcoming,bhatele2017massively}
perform some of the fastest epidemic simulations.
EpiSimdemics shows impressive scaling, presenting a zip-code
based partitioning scheme similar to the one employed by Loimos. They show
orders of magnitude difference in performance compared to previous work.
Table~\ref{tab:related} summarizes the performance of
the more performant models.

\section{Algorithm for Contagion Diffusion}
\label{sec:algorithm}
We develop our epidemic simulator, Loimos, using a combination of network
theory, discrete event simulation (DES), and agent-based modeling. We model both
individuals in the population and interactions between pairs of these \emph{agents}.
This allows us to simulate the dynamics of epidemic diffusion at a sufficiently
granular level to directly model the changing dynamics of disease spread in
the presence of a variety of public health intervention strategies.

\subsection{Serial Algorithm}
\label{sec:serial}

To expose parallelism across people and locations, we iterate over
discrete time steps. Each time step consists of:
\begin{enumerate}
  \item Identifying overlapping visits to each location.
  \item Calculating the likelihood that each overlap resulted in an infection
        and determining which infections occur.
  \item Updating each agent's disease state to reflect new infections and the
        progression of the disease.
\end{enumerate}

\vspace{0.08in}
\noindent\subsubsection{Disease Model and Finite State Automaton}
We assign each person, $p$, a disease state, $x_p$, managed using a finite state
automaton (FSA). The FSA specifies how people transition through various
disease states while infected, representing different stages of disease
progression. Each state has an associated susceptibility, $\sigma(x_p)$,
and infectivity, $\iota(x_p)$, which determines how they are treated in the
simulation; we consider $p$ to be \emph{susceptible} when $\sigma(x_p) > 0$
and \emph{infectious} when $\iota(x_p) > 0$. Transitions between states are
stochastic both in terms of the state transitioned to and how long a person
remains in a given state. Throughout this paper we use
an expanded version of the Susceptible, Exposed, Infectious, and Recovered
(SEIR) model~\cite{liu1987dynamical}.

\vspace{0.08in}
\noindent\textbf{Discrete Event Simulation:}
The discrete event simulation (DES) identifies which people are at a location,
$\ell$, at the same time and for how long, based on the set of visits
to~$\ell$ during a given simulation day. The DES splits each visit into two
events -- an arrival and a departure -- and orders them in a queue, $Q_\ell$,
based on when they occur. The DES then processes events from this queue, as shown
in Algorithm~\ref{algo:visits}, identifying all of the corresponding exposure events.

\begin{algorithm}[h]
  {\footnotesize
    \KwComputeExposures(event queue $Q_\ell$, disease states $X_\ell$):{\\
      \ForEach{\rm event $e \in Q_\ell$ \rm with corresponding visitor $p_e$}{
        \sethlcolor{soul_1}
        \lHlIf{\hl{$p_e$ \rm is susceptible}}{\hl{$V = V_s, V' = V_i$}}
        \lHlElseIf{\hl{$p_e$ \rm is infectious}}{\hl{$V = V_i, V' = V_s$}}
        \lHlElse{\hl{continue}}
        \eIf{$e$ \rm is an arrival} {
          \sethlcolor{soul_2}
          \hl{Add $p_e$ to visitor list $V$ corresponding to $p_e$'s disease state}\;
        } {
          \sethlcolor{soul_3}
          \hl{Remove $p_e$ from corresponding visitor list $V$}\;

          \sethlcolor{soul_4}
          \HlForEach{\hl{$p'$ \rm currently in the opposite visitor list $V'$}}{
            \hl{$c_\ell =$ }\hl{\KwContactProbability}\hl{($\ell$)}\;
            \HlIf{\hl{$p_e$ \rm and $p'$ \rm make contact with probability $c_\ell$}}{
              \sethlcolor{soul_5}
              \hl{Compute propensity of infection for $p_e$ and $p'$ during the
              period of co-occupancy $T$ using $X_\ell$}\;
              \lHlIf{\hl{$p_e$ \rm is susceptible}}{\hl{add exposure to list $E_{p_e}$}}
              \lHlElse{\hl{add exposure to list $E_{p'}$}}
            }
          }
        }
      }
    }
  }
  \caption{Computing exposures at a location, $\ell$}
  \label{algo:visits}
\end{algorithm}

While processing an event, $e$, we maintain lists of all susceptible and infectious people
currently at the location, $V_s$ and $V_i$ respectively. When processing an event,
we use the disease state, $x_{p_e}$, of the corresponding person, $p_e$, to select the
appropriate list, $V$, or skip them if they are immune or exposed
(\highlight{soul_1}{lines 3-5}). We add a person to $V$ when they arrive
(\highlight{soul_2}{line 7}) and remove them when they depart
(\highlight{soul_3}{line 9}). When a person leaves, we consider everyone in the
opposite list, $V'$, as potential contacts with a fixed probability, $c_\ell$, based on the
location (See~\eqref{eq:contact}; \highlight{soul_4}{lines 10-12}). If a contact occurs
between an infectious person and a susceptible person, we store the propensity of the
resulting exposure to cause an infection (See~\eqref{eq:propensity}; \highlight{soul_5}{lines 13-15}).

\vspace{0.08in}
\noindent\textbf{Contact Model:}
The contact model operates at each location, $\ell$, independently and
determines whether a pair of overlapping visits to $\ell$ result in a contact.
Here we adopt the $min/max/\alpha$ model formulation of Chen et
al.~\cite{chen2025epihiper}, which computes the
\emph{contact probability}, $c_\ell$, for any pair of people simultaneously
present at $\ell$ as a function of its maximum occupancy, $N_\ell$.
This serves as a proxy for its size. In order to compute $c_\ell$,
we select a minimum value, $A$, where if $N_\ell < A$
everyone will make contact, and a maximum value, $\alpha$, where if
$N_\ell > \alpha$ someone visiting at the peak occupancy of the location
will make about $B$ contacts; a person visiting a given location during peak
occupancy should expect to make between $A$ and $B$ contacts during that visit.
$c_\ell$ is given by,

\vspace{0.05in}
\begin{equation}
  \label{eq:contact}
  \eqnmarkbox[color_5]{c}{c_\ell} = \min\bigl\{1, \bigl[ \eqnmarkbox[color_1]{a1}{A}
    (\eqnmarkbox[color_2]{b1}{B} - \eqnmarkbox[color_1]{a2}{A})
    (1 - e^{-\eqnmarkbox[color_3]{n1}{N_\ell}/\eqnmarkbox[color_4]{al}{\alpha}})\bigr]
    / [\eqnmarkbox[color_3]{n2}{N_\ell} - 1]\bigr\}
\end{equation}
\annotate[yshift=.5em]{above}{c}{contact prob}
\annotatetwo[yshift=.5em]{above}{a1}{a2}{min contacts}
\annotate[yshift=-.25em]{below,left}{b1}{max contacts}
\annotate[yshift=-1em]{below, left}{n1,n2}{max occupancy}
\annotate[yshift=.5em]{above,right}{al}{threshold}
\vspace{0.1in}

\noindent for~\highlight{soul_3}{$N_\ell \ge 2$}.

We use the values \highlight{soul_1}{$A = 5$},~\highlight{soul_2}{$B = 40$}
and~\highlight{soul_4}{$\alpha=1000$} below, based on social
contact patterns in the POLYMOD data~\cite{mossong2008social}.

\vspace{0.08in}
\noindent\textbf{Transmission Model:}
The transmission model determines whether or not a given contact between
susceptible and infectious individuals ($p_i$ and $p_j$, respectively)
results in disease transmission. As in Chen et al.~\cite{chen2025epihiper},
transmission probabilities depend on a personal and a disease state susceptibility,
$\beta_\sigma(p_i)$ and $\sigma(x_{p_i})$ respectively, and infectivity,
$\beta_\iota(p_j)$ and $\iota(x_{p_j})$, the contact duration, $T$,
and a transmissibility, $\tau$.

The infection \emph{propensity}, $\rho$, of such a contact is given by,

\vspace{0.1in}
\begin{equation}
\label{eq:propensity}
  \begin{split}
    \eqnmarkbox[color_5]{r}{\rho}(p_i, p_j,
      \eqnmarkbox[color_4]{t1}{T}) = &
    \eqnmarkbox[color_4]{t2}{T}
    \cdot \eqnmarkbox[color_1]{tau}{\tau}
    \cdot \eqnmarkbox[color_3]{bs}{\beta_\sigma(p_i)}
    \cdot \eqnmarkbox[color_3]{s}{\sigma(x_{p_i})}
    \\
    & \cdot \eqnmarkbox[color_2]{bi}{\beta_\iota(p_j)}
    \cdot \eqnmarkbox[color_2]{i}{\iota(x_{p_j})}
  \end{split}
\end{equation}
\annotate[yshift=.5em]{above,left}{t2,t1}{overlap duration}
\annotate[yshift=-2.5em]{below,right}{r}{infection propensity}
\annotate[yshift=1.55em]{above,right}{tau}{transmissibility}
\annotate[yshift=.5em]{above,right}{bs,s}{susceptibility}
\annotate[yshift=-.5em]{below}{bi,i}{infectivity}
\vspace{0.1in}

\noindent where \highlight{soul_1}{$\tau$} is a tuning parameter proportional
to the likelihood of infection from a one-second contact with an infectious agent.

In order to determine whether or not a transition occurs for a given
susceptible person, $p_i$, at the end of a time step, we sum the propensities
from all $m$ contacts $p_i$ had with infectious people during the time step as shown below,

\vspace{0.1in}
\begin{equation}
   A(\eqnmarkbox[color_3]{pi1}{p_i}) = \sum_{j=0}^{m-1} \eqnmarkbox[color_5]{r}{\rho}(
    \eqnmarkbox[color_3]{pi2}{p_i},
    \eqnmarkbox[color_2]{pj}{p_j}, \eqnmarkbox[color_4]{t}{T_{i,j}})
\end{equation}
\annotate[yshift=.5em]{above}{r}{see \eqref{eq:propensity}}
\annotatetwo[yshift=-.75em]{below}{pi1}{pi2}{susceptible}
\annotate[yshift=-.75em]{below}{pj}{infectious}
\annotate[yshift=.5em]{above,right}{t}{overlap duration}
\vspace{0.1in}

\noindent where $T_{i,j}$ is the duration of the $p_i$'s $j$-th contact.
We then infect $p_i$ with probability $1 - e^{-A(p_i)}$.

\vspace{0.08in}
\noindent\textbf{Intervention Model:}
Interventions have three main components: a trigger, a selector, and an action.
The trigger activates the intervention at the end of a time step in response to
some global condition, such as the population passing a case threshold.  Once
the intervention is active, its selector determines which people or locations
it will apply its action to, based on their individual attributes and health
states.  This action will then either (1) add or remove some visits involving
the person or location in question or (2) adjust the values of attribute(s) of
the person or location, which may be either used in later interventions or
transmission propensity calculations. Most actions are designed to be reversed
once the intervention no longer applies to the person or location in question.

\subsection{Parallel Algorithm}
\label{sec:parallel}
The parallel algorithm, shown in Algorithm~\ref{algo:loimos}, operates on a
bipartite graph -- with people and locations as nodes, and weekly visit
schedules as edges -- along with an assignment of people and locations to
partitions. This algorithm is based on that originally proposed by Yeom et
al.~\cite{barrett2008episimdemics,yeom2014overcoming}.

\begin{algorithm}[t]
  {\footnotesize
  Partition $P$ into people partitions $\mathbb{P} = \{P_i\}$\;
  Partition $L$ into location partitions $\mathbb{L} = \{L_j\}$\;
  \ParFor{\rm person partition $P_i \in \mathbb{P}$}{
    \ForEach{\rm location partition $L_j \in \mathbb{L}$}{
      Compute the set $V_{i,j}$ of people on $P_i$ who visit some location on $L_j$\;
    }
  }

  \ForEach{\rm simulation day $d \in \{1, \dots, d_{\mathit{max}}\}$}{
    \sethlcolor{visits}
    \HlParFor{\hl{\rm person partition $P_i \in \mathbb{P}$}}{
      \HlForEach{\hl{\rm person $p \in V_{i,j}$}}{
        \hl{Send disease state update message $(p, x_p)$}\;
      }
    }

    \ParFor{\rm location partition $L_j \in \mathbb{L}$}{
      \HlForEach{\hl{\rm disease state update message $(p, x_p)$ to $L_j$}}{
        \hl{Store $p$'s disease state, $x_p$, in $X_\ell$}\;
      }
      \sethlcolor{inters}
      \HlForEach{\hl{\rm location $\ell$ \rm on $L_j$}}{
        \HlForEach{\hl{\rm visit $v$ \rm to $\ell$}}{
          \hl{Put an arrival and departure event into $Q_\ell$}\;
        }

        \hl{Reorder $Q_\ell$ by the time of event in ascending order}\;
        \hl{\KwComputeExposures}\hl{($Q_\ell, X_\ell$)}\;
        \HlForEach{\hl{\rm susceptible person $p$ \rm with exposure(s) at $\ell$}}{
          \hl{Send exposure message $m$ to $p$'s person partition}\;
        }
      }
    }

    \ParFor{\rm person partition $P_i \in \mathbb{P}$}{
      \ForEach{\rm person $p$ \rm on $P_i$}{
        \sethlcolor{inters}
        \HlForEach{\hl{\rm exposure message $m$ \rm destined for $p$}}{
          \hl{Put the exposures into the exposure list $E_p$}\;
        }

        \sethlcolor{eod}
        \HlIf{\hl{\KwIsInfected}\hl{($p, x_p, E_p$)}}{
          \hl{Update $p$'s disease state $x_p$}\;
        }
      }
    }
    Evaluate intervention triggers\;
  }
}
\caption{Parallel control flow in Loimos}
\label{algo:loimos}
\end{algorithm}

Prior to the main loop, we partition the person and location data ($P$ and $L$, respectively;
lines 1-2), and identify the subset of people, $V_{i,j}$, from each person
partition, $L_i$, that visit each location partition, $L_j$, (lines 3-7).

On each simulated day, each person partition sends a message with the current disease
state, $x_p$, of each person, $p$, it holds to each location partition its people visit
(\highlight{visits}{lines 9-13}). This update is performed in lieu of the full visit message
exchange from Yeom et al.~\cite{barrett2008episimdemics}, and is discussed in more depth in
Section~\ref{sec:loc-visits}. Once all the update messages have been received and the updated
disease state stored in $X_\ell$ (\highlight{visits}{lines 15-17}), arrival and departure events for each
visit are created and placed in a time-ordered queue, $Q_\ell$ (\highlight{inters}{lines 19-22}).
Locations currently selected by an active intervention may have altered visits. Each process performs a DES
for each of its locations as described in Algorithm~\ref{algo:visits}, determines whether each pair
of people whose visits overlap come into contact, and then computes the propensity
of any contacts based on their disease states. Once these calculations have been performed,
we send exposure messages to people with at least one exposure (\highlight{inters}{lines 24-26}).
Each person's exposure messages are then gathered in $E_p$
(\highlight{inters}{lines 31-33}) and processed to determine whether a given
exposed person was infected (\highlight{eod}{line 34}). If a susceptible
person is infected
or an infected person makes a timed transition, their disease state will be updated to reflect this at
the end of the simulation day (\highlight{eod}{line 35}). Finally, we evaluate the triggers of any interventions
deployed in the current scenario (line 39).

For simplicity, we primarily refer to the three phases of the main loop of this algorithm
in later performance analysis, namely the (1) person state communication
(\textbf{PSC}; \highlight{visits}{lines 9-17}), (2) exposure computation and communication
(\textbf{ECC}; \highlight{inters}{lines 18-33}), and (3) person state update
(\textbf{PSU}; \highlight{eod}{lines 34-36}).

\section{Implementation in Loimos}
\label{sec:impl}
We present a parallel epidemic modeling framework, Loimos, that implements
the parallel algorithm described in the previous section.  Loimos is written
on top of the Charm++ parallel runtime~\cite{CharmAppsBook:2013, kale:sc2011}.
In this section, we discuss salient details regarding the design of its parallel
implementation.

\subsection{Inputs to the Simulator}
\label{sec:inputs}

There are three core components that define an epidemic simulation:
\begin{enumerate}
  \item A population, consisting of people, locations, and visits.
  \item A disease, represented as a finite state automaton (see Section \ref{sec:serial}).
  \item An (optional) set of interventions, capable of modifying visit schedules and
  disease transmission likelihoods.
\end{enumerate}

We use two different types of populations for the simulations described in
this paper: realistic \emph{digital twin} populations mirroring
several U.S. states and purely synthetic populations.

\vspace{0.08in}
\noindent\textbf{Generating Realistic Populations:}
We generate these realistic datasets from a range of data sources through an
extension of the pipeline developed by Chen et al.~\cite{chen2025epihiper}.
For a given state, we begin by constructing a collection of people with demographics
(including age, gender, and occupation codes~\cite{census_north_nodate})
and partitioning these people into households. We refine these partitions
at a block group level, using iterative proportional fitting to match the demographic
distributions found in American Community Survey (ACS) data~\cite{census_census_nodate}.

Next, we assign each person a set of activities, using National Household Travel
Survey data through random forest methods conditioned on demographics and calibrated
against time-use surveys~\cite{fha_nhts_nodate}. We then construct home and activity
locations, integrating building~\cite{microsoft_microsoftusbuildingfootprints_2024}
and school~\cite{nces_electronic_nodate} data. Finally, we assign people to home
locations, and activities to activity locations. We constrain this assignment to
match ACS commute flow and demographic data (such as to ensure teachers work at schools).
We use these techniques to generate datasets representing all 50 U.S. states,
five of which are used in this paper (see Table~\ref{tab:strong-popsize}).
These datasets range in size from Arkansas to
California, with 2.75 and 35.5 million people, respectively,
allowing us to test strong scaling performance for a wide range of scales.

\begin{table}[h]
  \centering
  \caption{Realistic digital twin datasets used for strong scaling
  studies. Interaction and visit counts given per day.}
  \label{tab:strong-popsize}
  \begin{tabular}{lrrrr}
    \toprule
    Dataset Name & \# Interactions & \# Visits & \# People  & \# Locations \\
    \midrule
     Arkansas (AR)   & 63.65M       & 12.81M   & 2.749M & 13.17M \\
     Iowa (IA)       & 68.41M       & 14.24M   & 2.967M & 13.68M \\
     Michigan (MI)   & 226.5M       & 44.39M   & 9.342M & 16.33M \\
     New York (NY)   & 525.6M       & 88.28M   & 18.11M & 17.97M \\
     California (CA) & 955.7M       & 164.6M   & 35.51M & 24.49M \\
    \bottomrule
  \end{tabular}
\end{table}

\vspace{0.08in}
\noindent\textbf{Generating Purely Synthetic Populations:}
We generate our purely synthetic datasets on the fly, using a structured
grid of locations to maintain epidemic locality. Given an average number
of visits per person per day, $\lambda_\text{visits}$, we assign each person,
$p$, to a home location uniformly throughout the grid, then generate
$n \sim \text{Pois}(\lambda_\text{visits})$ visits for each person on each
day, where $\text{Pois}(\lambda)$ is the Poisson distribution with expected
rate $\lambda$. Here, the $i$-th visit is to a random location $d_i \sim
\text{Pois}(\lambda_\text{hops})$ hops away in the grid from $p$'s home location.
We use the average visits per person in the CA data, $\lambda_\text{visits}
= 4.6$, and $\lambda_\text{hops} = 5.2$.

Using the parameters above, we generate three series of datasets, each with different numbers of people and locations per core (as
shown in Table~\ref{tab:weak-popsize}). Each series (1x, 2x and 4x) can be used
to generate datasets on the fly at different core counts, thereby facilitating
weak scaling studies.

\begin{table}[ht]
  \centering
  \caption{Purely synthetic population datasets used for weak scaling.}
  \label{tab:weak-popsize}
  \begin{tabular}{rrr}
    \toprule
    Relative Size & \# People per core & \# Locations per core \\ \midrule
    1x            & 280k              & 70k                  \\
    2x            & 560k              & 140k                 \\
    4x            & 1.120M            & 280k                 \\
    \bottomrule
  \end{tabular}
\end{table}

\subsection{Task-based Runtime System}

Loimos is implemented in Charm++\cite{CharmAppsBook:2013, kale:sc2011}, a
parallel programming model and runtime system that provides asynchrony and
message-driven execution.  Charm++ has been used in production scientific
software such as NAMD~\cite{bhatele:ipdps2008, bhatele:ill2009},
OpenAtom~\cite{bhatele:europar2009}, and
EpiSimdemics~\cite{bhatele:ccgrid2017}.  When writing Charm++ programs, the
programmer organizes work and data into C++ objects called
\emph{chares}.  Chares are in turn organized into \emph{chare arrays}, which are indexable
collections of chares. At runtime, the Charm++ runtime system is responsible
for assigning chares to cores and for scheduling the execution of
chares assigned to a given core. Chares only start running when
a message is received for this chare from another chare. In Loimos, we use two chare
arrays: one for people and one for locations, with each chare containing a
partition of the appropriate data.

The other main Charm++ object we use in Loimos is a \emph{node group}.
Node groups have a single instance for each physical node the program is run on,
allowing us to avoid keeping redundant copies of shared information in
memory on one node, while also minimizing inter-node communication.

\subsection{Implementation of Different Models}

We implement the various models described in Section~\ref{sec:serial} modularly
to increase the flexibility of the codebase.

\vspace{0.08in}
\noindent\textbf{Disease Model:}
At the start of a simulation, we load the FSA representing the simulated
disease from an input file, which specifies its states and transitions.  A
Charm++ node group then stores a single copy of this information on each node,
in order to afford each person and location chare efficient read-only access
to these data. We store other read-only data describing the input scenario in a
similar fashion.

When initializing the simulation, people begin in one of several
entry disease states, as determined by their individual attributes, such as age,
specified in the input file. They remain in this state until they are infected
by an infectious person or chosen to seed the outbreak. For this work, we select
a small sample of people to infect during the first few days of the simulation.
For the runs in this paper, we infect two people per day, chosen uniformly at
random, for the first ten days of the simulation, and use an FSA representing
COVID-19 with five  age-based entry states, each with 18 distinct reachable states.

\vspace{0.08in}
\noindent\textbf{Discrete Event Simulation:}
On each simulation day, we start the DES after the conclusion of the person state
communication (PSC), when updates to interventions (and thus visit schedules) are
also guaranteed to have completed. We use Charm++'s quiescence detection
mechanism
-- a soft barrier which ensures no messages are in flight or being processed
before continuing -- to determine when this phase is done. This approach
allows Loimos to detect the end of the phase even though the number of visits
a given location will receive is both not known
\emph{a priori} and non-deterministic in the presence of any intervention
which changes visit schedules, such as school closures. Once triggered,
each location chare independently executes the DES for its assigned locations.

In implementing the DES algorithm, we made three key optimizations: (1) we only
keep track of co-occupancy, and thus interactions, between susceptible
people and infectious people, (2) we only send
exposure messages to susceptible people who experienced at least one exposure
during a time step, and (3) exposure messages are
sent as soon as we finish processing a susceptible individual's departure
event.  Note that we are able to make the first optimization without affecting the
results of the simulation because only contact between a susceptible person and
an infectious person constitutes an exposure (and thus can cause an infection).
The second optimization is especially helpful as in most iterations only a small
fraction of visits will result in an exposure.  The third optimization allows us
to significantly overlap computation and communication during this phase.

\vspace{0.08in}
\noindent\textbf{Contact Model:}
We implement two different types of contact models in our code. The first is the
$min/max/\alpha$ model (see Section~\ref{sec:serial}). Since this requires
knowledge of the maximum occupancy of each location, we use a pre-processing
script to compute this based on the visit schedules file and save each
location's maximum occupancy to the locations file. At the start of a run,
we read in this value for each location, compute the appropriate contact
probability, and store it as a new location attribute. Since the maximum occupancy cannot be
computed in advance for the purely synthetic datasets, we also provide a second
contact model with a single, global contact probability.

\vspace{0.08in}
\noindent\textbf{Transmission Model:}
After each exposure is identified, we compute the corresponding propensity
and batch it with the other exposure messages for the
susceptible person involved. We send the exposure messages for each person
immediately after processing the departure event for their visit in the DES.
We do not compute actual infections until after the DES is complete and all
exposure messages have been received. Again, we use Charm++'s quiescence detection to
determine when this has occurred, as the number of exposure messages each
person chare will receive is non-deterministic, depending on the mixture
of infectious and susceptible people as well as the contact model. Once all messages
are received, each person chare sums the propensities for each
person in their partition in order to identify infections.

\vspace{0.08in}
\noindent\textbf{Intervention Model:}
Similar to the disease model, the specifications for the interventions to be
used in a given run are provided in an input file and stored on a node group.
Unlike the disease model, we store and update some state for each intervention
via the main chare over the course of the simulation. In particular, at the end
of each simulation day, we perform a reduction across all person chares to
compute the number of total infectious people, and pass this value along with
the current day to the triggers for each specified intervention to determine
which interventions should be active. The IDs for each active location-based
intervention are then passed to all location chares. These chares then access
the local copy of the intervention objects on their node and filter the
locations assigned to that chare using the selector for the intervention, and
apply the action to the relevant locations.

We use a separate class for each intervention, each of which extends a shared
interface with methods for testing whether a location should be selected,
making arbitrary changes to a location's state via an action (including
changing its visit schedule), and undoing the changes wrought by the action.
Some interventions such as vaccination have a trivial undo method as the changes
persist after the initial intervention ends. We use a similar scheme for
person-based interventions.

\section{Additional Performance Considerations}
\label{sec:opt}
\begin{figure*}[t]
    \centering
    \includegraphics[height=1.95in]{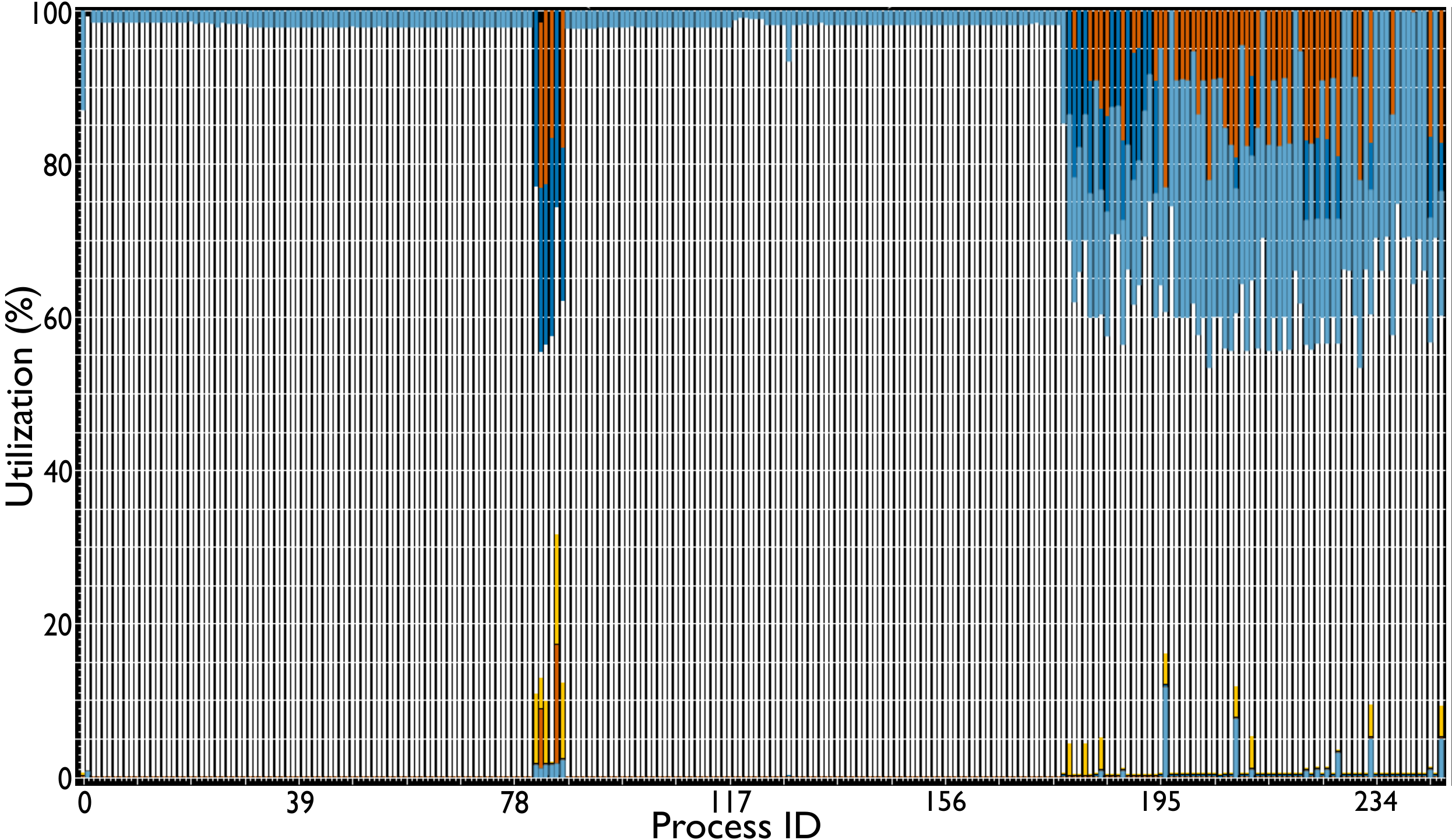}
    \hfill
    \includegraphics[height=1.95in]{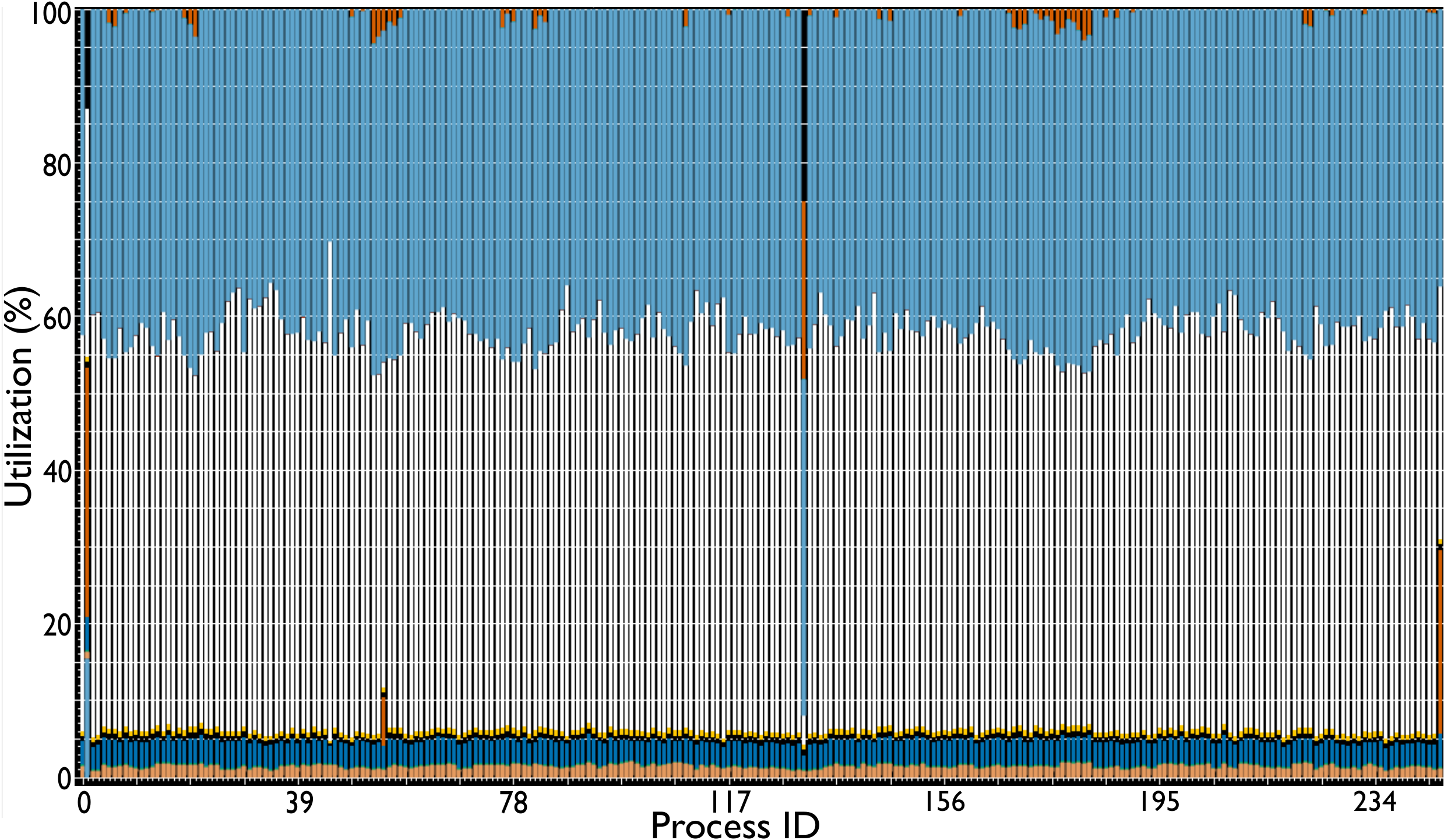}
    \includegraphics[width=\linewidth]{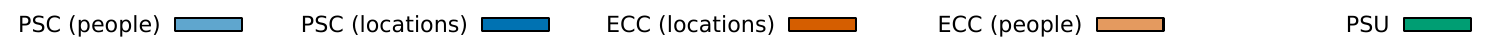}
    \caption{Visualizations generated using the Projections tool showing
    processor utilization in three iterations of Loimos on two Perlmutter
    nodes (256 cores) with MI data. Note that there is much more idle time
    on a larger number of cores with no optimizations (left),
    than with static load balancing only (right).}
    \label{fig:opts-lb}
\end{figure*}

After implementing Loimos's core features, we explored several avenues to
optimize its performance further. The four directions we found most
beneficial are: (1) considering different combinations of processes and threads
per node, (2) static load balancing, (3) short-circuit evaluation of the DES at
each location, and (4) storing visit data on location chares.

\subsection{Impact of Using Processes vs.~Threads}
\label{sec:smp}

Charm++ provides support for different machine layers as well as
abstractions that enable the programmer to adapt to the underlying hardware
to improve performance and scalability. All these components make Charm++
codes highly tunable. In particular, we are interested in analyzing how running
Loimos with symmetric multiprocessing (SMP) support in Charm++ performs relative to
the non-SMP alternative.

Enabling the Charm++ SMP mode is analogous to adding OpenMP or another threaded
programming model to an MPI code, in that it runs multiple
threads per process instead of the usual single thread. Users can specify
how many worker threads are spawned per process,
along with an optional mapping from threads to cores, but one thread per
process is always dedicated to communication. This communication thread manages
inter-node messages whereas intra-node communication is managed through the
shared memory address space common to all threads on the same node. The
requirement of allocating one communication thread per process could be a
disadvantage for compute-intensive applications since compute cores have to
be sacrificed. However, for communication-intensive applications, the use of a
dedicated communication thread to manage messaging might lead to better
performance, or lead to poor performance by becoming a bottleneck.

We compare five different ratios of processes to nodes,
using 8 processes per node (p/n) with 15 or 14 worker threads per process (t/p)
and 16 p/n with 7 or 6 t/p in SMP mode, and 126 processes per node in non-SMP mode.
We performed a strong scaling experiment, running three replicates of each
configuration for 200 days on the MI data (see Table~\ref{tab:strong-popsize})
with all optimizations enabled. All scaling experiments
in this paper were conducted on the Perlmutter Cluster at NERSC, a HPE Cray EX cluster
with two AMD EPYC 7763 CPUs per node, each with 64 cores, and a HPE Slingshot 11
interconnect~\cite{perlmutter}.

\begin{figure}[h]
  \centering
  \includegraphics[width=\columnwidth]{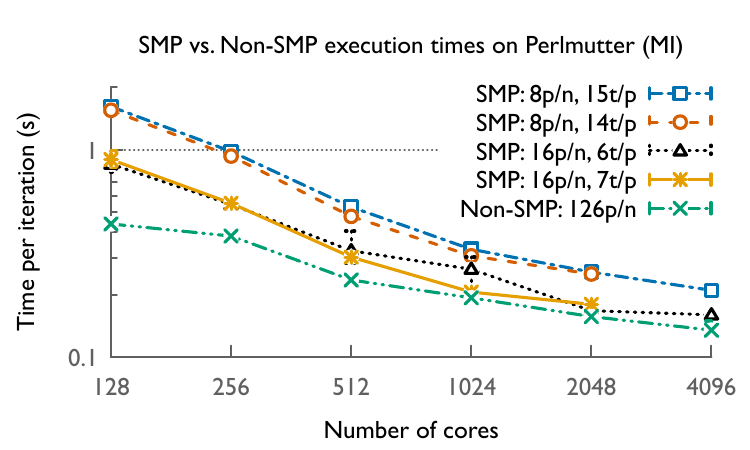}
  \caption{Strong scaling comparison of the performance of different symmetric
    multiprocessing (SMP) configurations on Perlmutter with MI data.
    All four SMP configurations
    -- with different processes per node (p/n) and worker threads per process
    (t/p) counts -- perform worse than the non-SMP configuration for all
    core counts. Execution times are averaged over three runs, with
    extrema shown in error bars.}
  \label{fig:smpconfigurations}
\end{figure}

Figure~\ref{fig:smpconfigurations} shows the experimental results of this SMP
vs. non-SMP comparison. The non-SMP configuration
consistently out-performs the SMP configurations, achieving a speedup of
about $3.25 \times$ when going from 128 cores to 4096. The 16 p/n SMP
configurations similarly out perform the 8 p/n SMP ones, perhaps due to their
usage of additional communication threads. Note that the SMP
configurations suffer from fatal runtime errors on larger core counts.
On 4096 cores, the 126 p/n configuration is $1.19 \times$ faster
than the best SMP configuration (16 p/n with 6 t/p). As a result, we use the
126 p/n non-SMP configuration for all runs described in subsequent sections.

\begin{figure*}[t]
    \centering
    \includegraphics[height=1.5in]{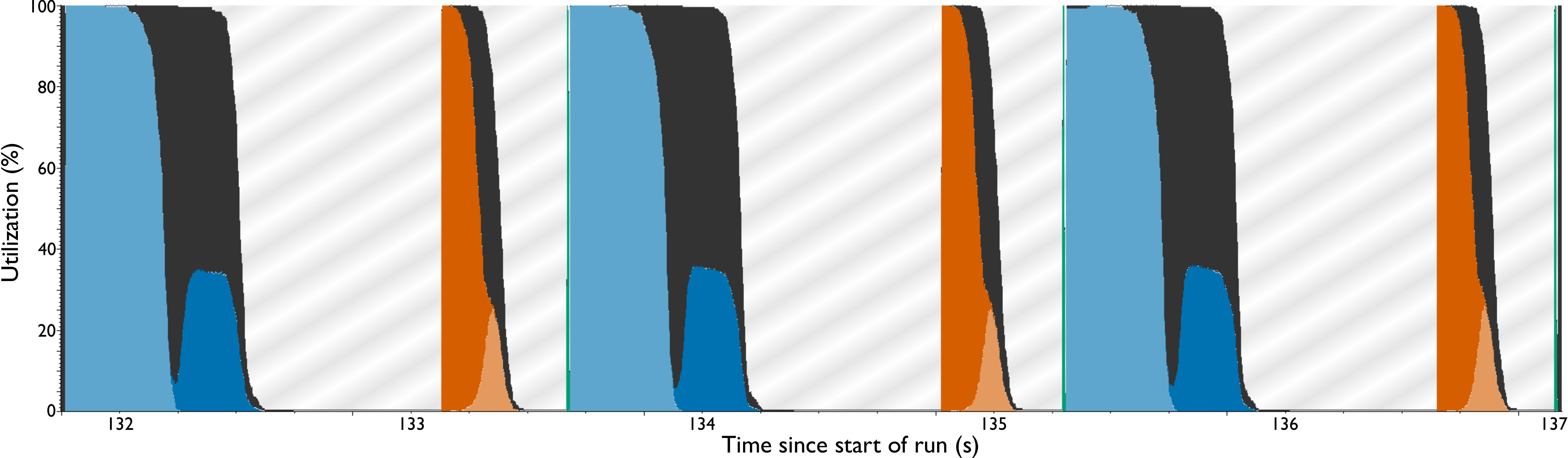}
    \hfill
    \includegraphics[height=1.5in]{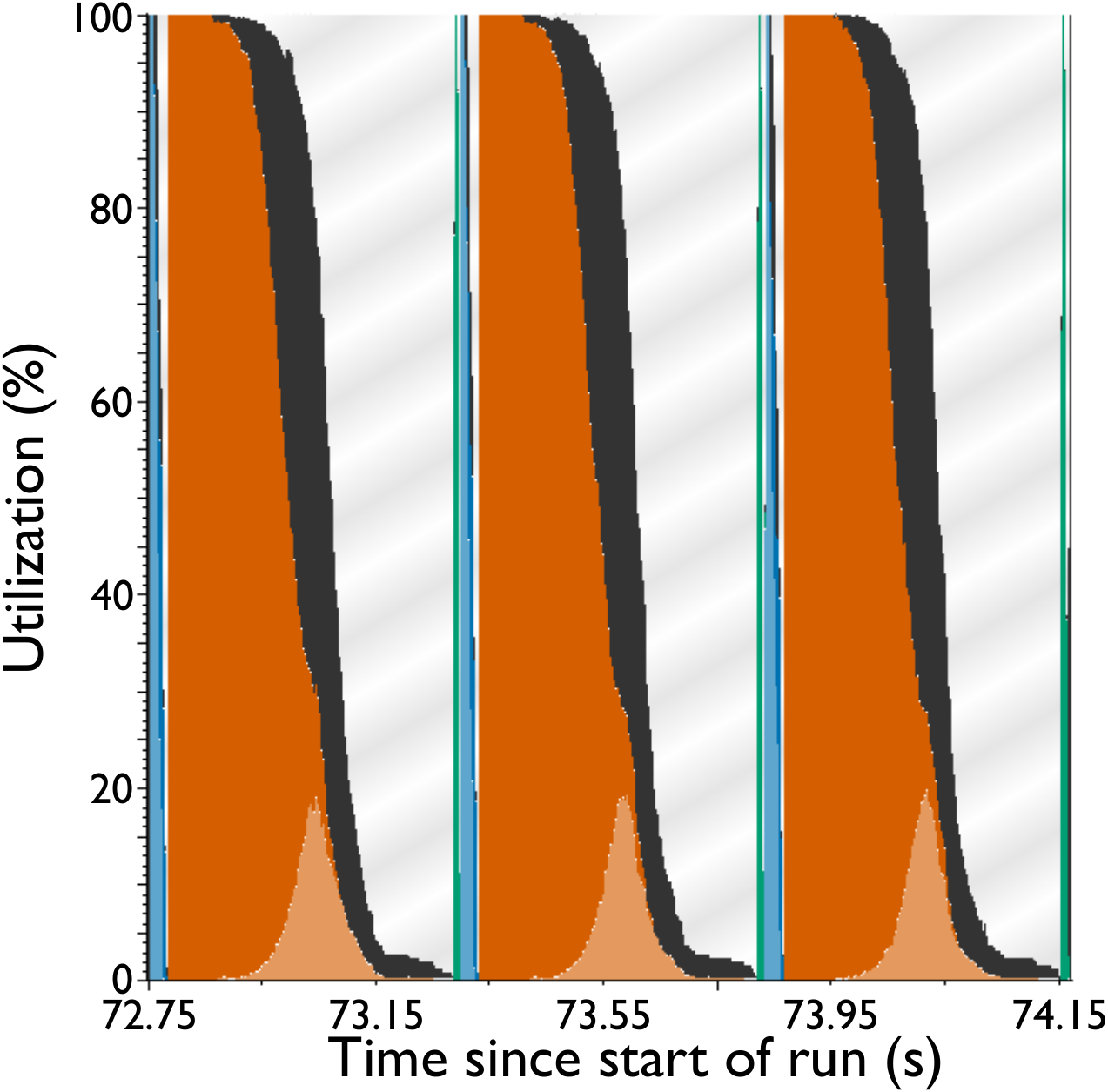}
    \includegraphics[width=\linewidth]{legend.pdf}
    \caption{Visualizations generated using the Projections tool showing the
    breakdown of time spent in three iterations of Loimos on two
    Perlmutter nodes (256 cores) with MI data.
    We observe that most of the time is spent in the person state
    communication (PSC) phase with only static load balancing and short
    circuit evaluation of interactions (left), but negligible time doing so
    when storing visits on location chares (right). In the latter case, the
    time spent in the exposure computation and communication phase (ECC)
    dominates, and the total execution time is significantly reduced.}
    \label{fig:opts-loc-visits}
\end{figure*}

\subsection{Static Load Balancing}
\label{sec:lb}

During initial runs of Loimos on the realistic datasets, profiles collected
using the Projections tool~\cite{namdPerfStudy} revealed that some
processes ran much slower than others, as shown in the left of
Figure~\ref{fig:opts-lb}. Projections profiles break down a program's runtime
by Charm++ tasks; in Loimos, these tasks correspond to the portion of
each simulation phase (PSC, ECC, or PSU) which happens on either people or
locations chares, plus idle time (white) and overhead (black) across the
whole program. We set out to minimize the
load imbalance shown in Figure~\ref{fig:opts-lb} by improving the assignment
of locations and people to chares.

Toward this end, we present a simple static partitioning scheme meant to
preserve geographical locality. First, we sort all locations in the population
by the ID of their state, county, census tract, and census block group, in that
order. This is intended to ensure that nearby locations are placed on the same
chare if possible. Using the number of visits to a location, $\lambda_j$, as a
proxy for its load, we compute the average load per chare, $\Lambda$, as the
ratio of visits to location chares, for a given number of chares. If any
locations have a load $\lambda_j > \Lambda$, we assign them their own chare
and recompute $\Lambda$ for the remaining chares, until no such locations
remain. We then compute the cumulative load for each location, fixing
$\lambda_j = \Lambda$ for those heavy locations already assigned to a
chare, and assign each location to chare $\left \lfloor \left (
\sum_{i=0}^{j-1} \lambda_i \right ) / \Lambda \right \rfloor$.

We then partition the person data, identifying the location chare, $L_j$, containing
each person's home location and placing that person on person chare $P_j$. Applying this scheme
drastically decreases the idle time on most processes, as shown in the right of
Figure~\ref{fig:opts-lb}, although a much lower degree of imbalance still
remains. Figure~\ref{fig:opts} shows how Loimos's performance improves
(no-opts vs. static) when using this load balancing scheme. This leads to a
speedup of $5.55 \times$ and $3.55 \times$ for the MI data on 128 and 4096
cores, respectively.

\subsection{Optimizing the Interaction Computation}
\label{sec:sc}
\begin{figure}[h]
    \centering
    \includegraphics[width=\columnwidth]{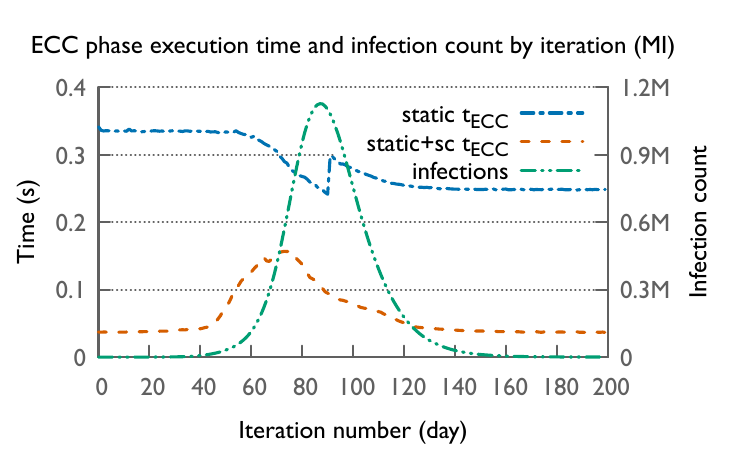}
    \caption{A 200-day simulation of Loimos on two Perlmutter nodes (256 cores)
    with MI data. We observe that time spent on exposure computation and
    communication is much greater without (static $t_\text{ECC}$) than with
    (static+sc $t_\text{ECC}$) short circuit evaluation of interactions.
    In the latter case, execution time is highest when cases (infections) are
    increasing fastest.}
    \label{fig:opts-sc}
\end{figure}

Our next optimization was inspired by observing significant variation in the time we spent in one
of the three main simulation phases over the course of a run. We
initially expected the time spent in the exposure computation and communication (ECC)
phase to roughly track the number of infections, due to our use of separate queues
for infectious and susceptible visitors.
Instead, we observed that ECC dominates the runtime for the first half of the
simulation, but falls after the peak of the infection curve has passed,
as shown in Figure~\ref{fig:opts-sc}. This corresponds to an increasing
number of immune people in the population, who are ignored when identifying
exposures.

As a result, we realized that we spend significant time computing
exposures for every location even when no infectious individuals are
present, as we still process all susceptible arrival and departure events in
the DES. In order to avoid this, we add a check to skip the DES
entirely on locations with no infectious people visiting them on a
given day. As shown in Figure~\ref{fig:opts-sc}, after implementing this
\say{short-circuit} computation of the DES, the ECC runtime corresponds more closely to
the number of infectious people, peaking near the first inflection point of
the epidemic curve. Figure~\ref{fig:opts} shows the benefits of using this
scheme (static+sc) on top of the previous optimization (static),
which results in about a $1.67 \times$ and $1.21 \times$ speedup for the MI
data on 128 and 4096 cores.

\subsection{Storing Visits on Location Chares}
\label{sec:loc-visits}

Finally, Projections~\cite{namdPerfStudy} profiles revealed that sending visit
data from person to location chares (PSC), which initially began each
iteration, dominated the simulation runtime, as shown in the
left of Figure~\ref{fig:opts-loc-visits}.
This was due to the fact that the visit data was stored on person chares, as in
Yeom et al.~\cite{yeom:ipdps2014}, despite primarily being used in the DES
performed on location chares. This exchange also resulted in substantial idle
time (white) before the start of the next phase (ECC).

\begin{figure}[h]
  \centering
  \includegraphics[width=\columnwidth]{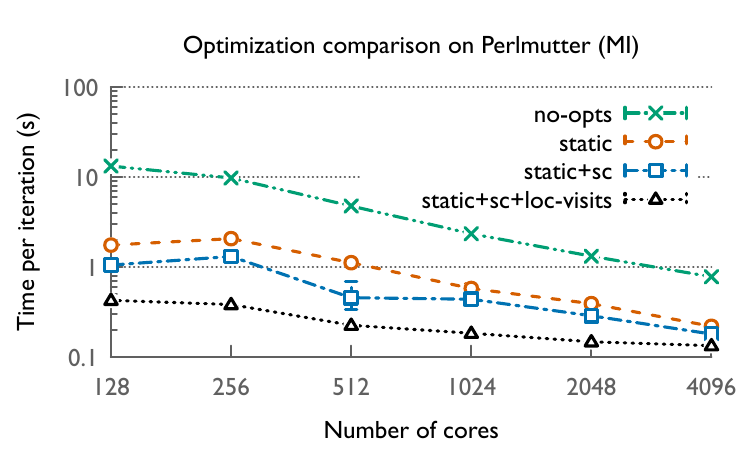}
  \caption{Performance impact of adding the following optimizations over
  the original Loimos implementation (no-opts):
  (1) static load balancing (static), (2) short circuit evaluation of
  interactions (sc), and (3) storing visit data on location chares (loc-visits).
  Each added optimization reduces runtimes, which are averaged over
  three runs on the MI data on Perlmutter with extrema shown in error bars.
  }
  \label{fig:opts}
\end{figure}

We hypothesized that we could reduce the amount of
communication by storing the visit data where it is used and only communicating
data that changes between iterations, namely updated disease states and modified
per-person susceptibility and infectivity values. Visit schedule changes resulting
from interventions could then be evaluated on location chares based on a cache containing
the relevant person state data. After implementing this change, we found we spend minimal time in the new
person state communication (PSC) phase, as shown in the right of Figure~\ref{fig:opts-loc-visits},
which previously dominated simulation runtimes. As a side effect, however, we spend more
time evaluating the DES, as we can no longer overlap the queuing of arrival and departure
events with communicating visit data. Figure~\ref{fig:opts} shows the overall effect
of this optimization (static+sc+loc-visits). This results in a $2.46 \times$
and $1.34 \times$ speedup on 128 and 4096 cores.

\begin{figure*}[t]
    \centering
     \includegraphics[width=\columnwidth]{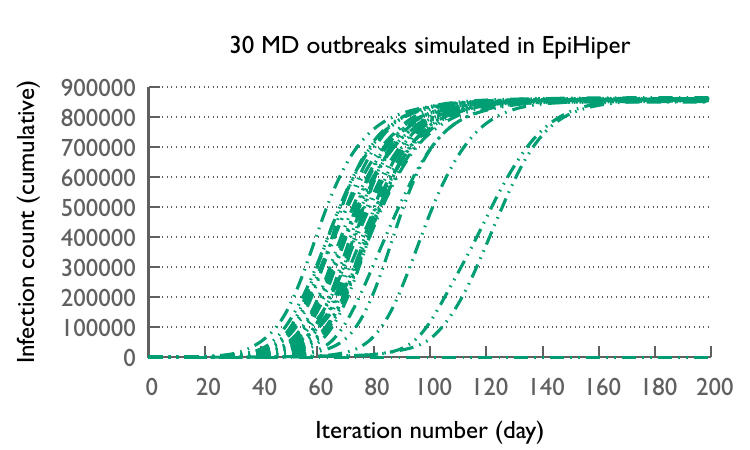}
     \includegraphics[width=\columnwidth]{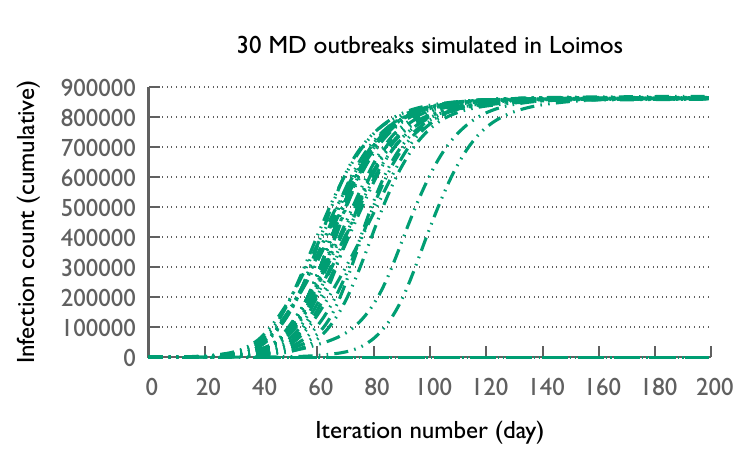}
    \caption{Cumulative infections over time for 30 replicates in EpiHiper (left)
    and Loimos (right) of a simulated MD outbreak, with both distributions showing
    similar average infection totals but with a wider range of times until equilibrium is reached
    for EpiHiper.}
    \label{fig:validation}
\end{figure*}

\begin{figure*}[t]
    \centering
     \includegraphics[width=\columnwidth]{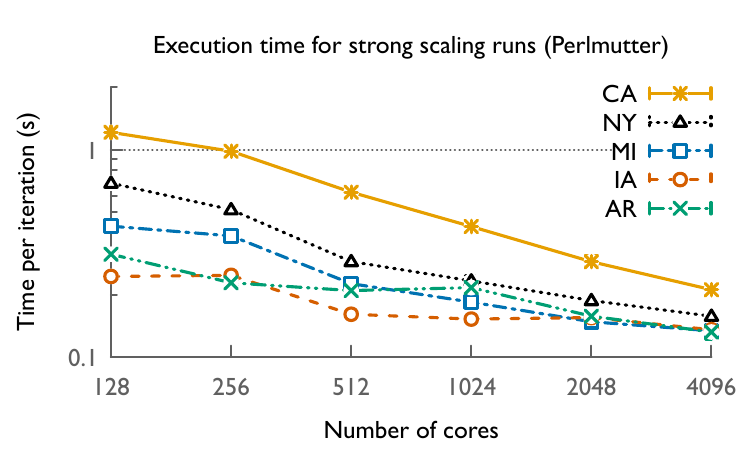}
     \includegraphics[width=\columnwidth]{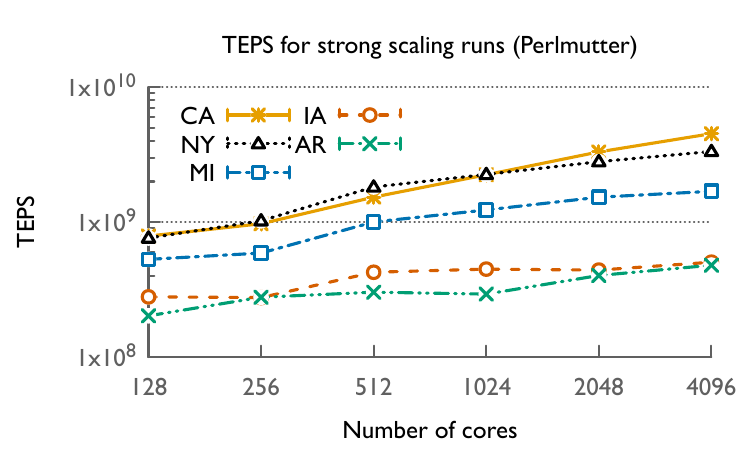}
     \caption{Strong scaling performance of Loimos on Perlmutter~for five different datasets
     in terms of execution time (left) and traversed edges per second (TEPS, right).
     Both show modest, but consistent, linear speedups for larger datasets.
     Execution times averaged over three runs, with extrema shown in error bars.}
     \label{fig:strong}
\end{figure*}

When all optimizations are combined together, we achieve a significant $31.03 \times$ speedup
on 128 cores and a $5.83 \times$ speedup on 4096 cores compared to the baseline with no optimizations.

\section{Experimental Setup}
\label{sec:setup}
We began by performing a distributional docking study
(see~\cite{collins2024methods}) to validate Loimos's simulation output against
that of an established epidemiological simulation,
EpiHiper~\cite{machi_scalable_2021}. For this study, we used a dataset with
5.513 million people and 2.896 million locations representing Maryland (MD)
for which we developed versions that can run in both models. This dataset was
developed using the same methods as those shown in Table~\ref{tab:strong-popsize}.

For 30 runs, we varied the random seed to capture
the distribution of potential epidemiological outcomes. These results were
then compared against runs of the existing EpiHiper simulation using the same
input visit network and simple SIR disease model. Note that EpiHiper used the
visit data differently than Loimos: a preprocessing script determined the list
and duration of contacts for each person in the population, producing a fixed
contact network which the disease then diffused over. Each EpiHiper run was on
a separate contact network. In contrast, Loimos determined a person's contacts
separately on each day, effectively resulting in a dynamic contact network,
even in the absence of interventions. For all validation runs, the
transmissibility was fixed at $\tau=0.05$.

We then performed extensive scalability studies using the 126 cores per
node non-SMP configuration (See Section~\ref{sec:smp}) on Perlmutter.
All scaling experiments were run with Protobuf version 3.21.12 and Charm++
version 7.0.0. All scaling runs used the same random seed, and thus had
identical epidemiological results. Values shown were the average of three
replicates, with the error bars representing the minimum and maximum runtimes.
We used the realistic datasets shown in Table~\ref{tab:strong-popsize} for
strong scaling and those in Table~\ref{tab:weak-popsize} for weak scaling.
See Section~\ref{sec:impl} for a more detailed description.

Table~\ref{tab:strong-popsize} describes the number of people, locations, and
total visits for all these datasets. Note that interactions (person-person
edges) are given on average, due to the stochasticity of the contact model. We
ran the realistic datasets for 200 days
total, processing a total of about 21 and 191 billion
interactions for the AR and CA datasets, respectively, over the course of the simulation.
In order to ensure a representative workload, the transmissibility of the simulated
outbreaks was tuned so that the number of infectious people peaked about
halfway through the simulations.

Next, we performed a weak-scaling experiment. We ran three fixed
problem sizes per process, as shown in Table~\ref{tab:weak-popsize}.
These datasets were generated using the on-the-fly synthetic population generation
method outlined in Section~\ref{sec:impl}.

We evaluated the performance of all scaling runs by calculating the average
execution time per simulation day, excluding data loading and application
startup time.

\section{Validation of Loimos}
We first present the results of validating Loimos against
EpiHiper~\cite{machi_scalable_2021}. We sought to show that the
distribution of Loimos's results corresponds to those of an established simulator,
EpiHiper, with a focus on total cumulative infections and the time to reach an equilibrium
state. Figure~\ref{fig:validation} illustrates how both simulators show similar
overall disease trajectories, with outbreaks either persisting to infect a
significant proportion of the population or dying out quickly.  In the former
case, both simulations average similar numbers of total cumulative infections
-- 863k for Loimos and 858k for EpiHiper -- and the latter outcome occurs rarely
for both simulations -- twice for Loimos and once for EpiHiper.

With respect to the time to equilibrium for the persistent outbreaks, Loimos shows more tightly
clustered results than EpiHiper.  This is likely a byproduct of how the two
simulators handle their input networks. Since EpiHiper uses the same contact
network for an entire run, differences in the chosen contact network have the
potential to cause compounding differences in the simulation results. In
contrast, since Loimos essentially selects a new contact network in each
iteration, differences in contact networks between runs tend to be smoothed
over to some extent as more networks are sampled over the course of a run,
similar to how there is less variation in the average of 100 die rolls than
that of a single roll.

\section{Performance Results}
\label{sec:results}
We now present scaling results from benchmarking Loimos using various input
datasets on Perlmutter.

\subsection{Strong Scaling Performance}
\label{sec:strong}

In order to understand how Loimos would enable large scale simulations, we start
with performing strong scaling studies as shown in Figure~\ref{fig:strong}.
For reference, EpiSimdemics takes 2.67 seconds per day on 192 cores of Blue Waters
to simulate their California population~\cite{yeom2014overcoming},
whereas Loimos takes 1.21 seconds per day on 128 cores.
All five datasets achieve their best performance in terms of the average time per step
on 4096 cores (left plot). Notably, the smallest states scale inconsistently, with performance
remaining roughly flat from 256 to 1024 and 512 to 2048 cores for the AR and IA datasets,
respectively. The larger datasets, however, display consistent, if modest, linear
scaling. In terms of traversed edges per second (TEPS, right plot), Loimos
achieves the best TEPS for the NY dataset on up to 1024 cores. Beyond that
point, it performs best on the CA dataset, peaking at about 4.6 billion TEPS
on 4096 cores.

\begin{figure}[h]
    \centering
    \includegraphics[width=\columnwidth]{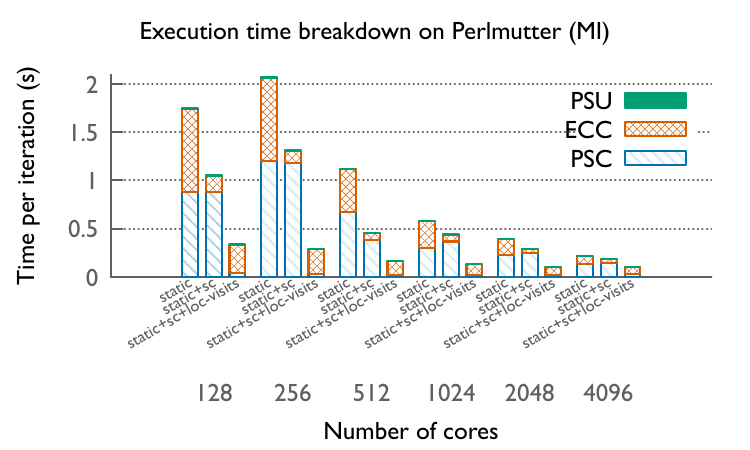}
    \caption{Breakdown of total time spent in Loimos into three phases:
    the (1) person state update (\hleod{PSU}), (2) exposure computation
    and communication (\hlinters{ECC}), and (3) person state communication
    (\hlvisits{PSC}) phases. Reductions in execution times are shown as
    three optimizations are incrementally applied: (1) static load balancing
    (static), (2) short circuit interaction computation (sc), and (3) location
    chare visit data storage (loc-visits).}
    \label{fig:scaling-breakdown}
\end{figure}

Figure~\ref{fig:scaling-breakdown} shows the breakdown of total time spent in
Loimos into the three simulation phases identified in Algorithm~\ref{algo:loimos}:
(1) the person state communication (PSC, \highlight{visits}{lines 6-15}),
(2) exposure computation and communication (ECC, \highlight{inters}{lines 17-27}),
and (3) the final person state updates (PSU, \highlight{eod}{lines 29-30}),
when the different optimizations are applied.
We observe that without the short circuit interaction optimization (static),
PSC takes slightly more time than ECC on all core counts, with the difference
growing as the core count increases, whereas with that optimization (static+sc),
ECC consistently takes a fraction of the time of PSC. When we store visit data
on location chares (static+sc+loc-visits), PSC takes negligible time, and ECC
is somewhat slower -- as queuing arrival and departure events can no longer be
overlapped with PSC. In all cases, the time spent in PSU is negligible.

\subsection{Weak Scaling Performance}
\label{sec:weak}

\begin{figure}[h]
\centering
 \includegraphics[width=\columnwidth]{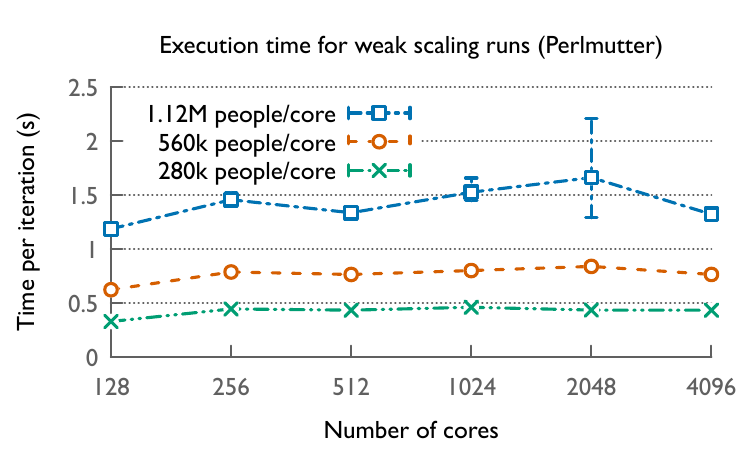}
 \caption{Weak scaling results on Perlmutter for three different synthetic datasets,
 showing relatively flat runtimes with increased variation for larger datasets.
 Execution times are averaged over three runs, with extrema shown in error bars.}
 \label{fig:weak}
\end{figure}

We also perform weak scaling tests to see how well Loimos handles datasets of
increasing size (from Table~\ref{tab:weak-popsize}). Figure~\ref{fig:weak} displays Loimos's relatively flat weak
scaling performance up to 4096 cores. For all configurations, there is a noticeable
increase in runtime from 128 to 256 cores, though this change is relatively
small -- representing a 35\%, 26\%, and 23\% slowdown for the 280k, 560k, and
1.12M people per core configurations, respectively. The latter
configuration shows increased variability on 1024 and 2048 cores along with
increased runtime, though both runtime and variability fall on 4096 cores.

Although these scaling results represent an optimistic representation of the
person-location visit graph, they exploit location-load attributes that are
present in the more realistic networks. While our simulation would see
significant slowdowns from purely random visits, our static load balancing
scheme is designed to co-locate people and locations and maximize
the interconnectedness between these objects.

\section{Conclusion}
\label{sec:conclusion}
Uncontrolled spread of infectious disease is a challenging societal issue --
one that requires policy makers to have the best possible tools in order to
make informed decisions. Computer simulations are one such vital tool. The
tight time constraints on relevant policy decisions mean that these simulations
need to be able to model large regions extremely quickly and accurately across
a wide variety of counterfactual scenarios. These demands require the use of
powerful supercomputing systems. Toward this end, we presented a scalable parallel
simulation framework for modeling contagion processes, Loimos, and demonstrated
its capabilities.

In this work, we outlined the methods we used to develop this simulation
framework and to optimize it for production HPC systems. We described the
models underpinning our work as well as various optimizations we have made to
enable the code to scale well. We demonstrated our code's use of resources
during both strong and weak scaling runs on Perlmutter at NERSC, achieving
modest, but linear, strong scaling speedups and relatively flat weak scaling
results. We also showed how the epidemiological results of the simulation compare
to an existing model. Together, these runs demonstrate the potential uses of
Loimos for policymakers as a fast epidemic simulator that is robust enough to
capture the effects of different policy interventions.

\section*{Acknowledgment}
This material is based upon work supported in part by the U.S.~Department of
Energy, Office of Science, Office of Advanced Scientific Computing Research,
Department of Energy Computational Science Graduate Fellowship under Award
No.~DE-SC0021.  It is also based upon work supported in part by the National
Science Foundation under Grant No.~CCF-1918656 and CINES-191680, and by the
Department of Defense, Defense Threat Reduction Agency, under Grant
No.~HDTRA1-24-R-0028.

This research used resources of the National Energy Research Scientific
Computing Center (NERSC), a U.S.~Department of Energy Office of Science User
Facility located at Lawrence Berkeley National Laboratory, operated under
Contract No.~DE-AC02-05CH11231 using NERSC awards DDR-ERCAP0032257 and
DDR-ERCAP0029890. The authors acknowledge Research Computing at the University
of Virginia for providing computational resources that have contributed to the
results reported within this publication.

\IEEEtriggeratref{41}
\bibliographystyle{IEEEtran}
\bibliography{cite,pssg}

\end{document}